# Ni cluster embedded (111)NiO layers grown on (0001)GaN films using pulsed laser deposition technique

Simran Arora[1], Shivesh Yadav[2], Amandeep Kaur[1], Bhabani Prasad Sahu[1], Zainab Hussain[1], Subhabrata Dhar[1*]

[1]Department of Physics, Indian Institute of Technology Bombay, Mumbai - 400076, India

[2]Department of Condensed Matter Physics and Material Science, Tata Institute of Fundamental Research, Mumbai - 400005, India

*E-mail: dhar@phy.iitb.ac.in

**Abstract**

(111) NiO epitaxial layers embedded with crystallographically oriented Ni-clusters are grown on c-GaN/Sapphire templates using pulsed laser deposition technique. Structural and magnetic properties of the films are examined by a variety of techniques including high resolution x-ray diffraction, precession-electron diffraction and superconducting quantum interference device magnetometry. The study reveals that the inclusion, orientation, shape, size, density and magnetic properties of these clusters depend strongly on the growth temperature ($T_G$). Though, most of the Ni-clusters are found to be crystallographically aligned with the NiO matrix with Ni(111)∥NiO(111), clusters with other orientations also exist, especially in samples grown at lower temperatures. Average size and density of the clusters increase with $T_G$. Proportion of the Ni(111)∥NiO(111) oriented clusters also improves as $T_G$ is increased. All cluster embedded films show ferromagnetic behaviour even at room temperature. Easy-axis is found to be oriented in the layer plane in samples grown at relatively lower temperatures. However, it turns perpendicular to the layer plane for samples grown at sufficiently high temperatures. This reversal of easy-axis has been attributed to the size dependent competition between the shape, magnetoelastic and the surface anisotropies of the clusters. This composite material thus has great potential to serve as spin-injector and spin-storage medium in GaN based spintronics of the future.

GaN, a wide and direct band gap (3.45 eV) semiconductor with wurtzite crystal structure is the material for display and lighting devices of modern era. It also has tremendous prospect in high frequency and high-power electronics.[1–4] Due to weak spin-orbit coupling, spin-flip scattering rate in GaN is less than that of other important semiconductors such as Si and GaAs, which makes this material a good choice for the spin flow medium in a spin transistor.[5–7] However, the biggest challenge for the realization of spin-transistor is to find a suitable spin-injector. Dilute magnetic semiconductors (DMS) and ferromagnetic semiconductors are preferred over metallic ferromagnets as spin-injecting materials for a semiconductor. Even though, metallic ferromagnets can offer high Curie temperature, efficiency of spin injection is weakened due to the mismatch in electrical conductivity between metal and semiconductor.[8,9] On the other hand, DMS/ferromagnetic semiconductors typically offer low Curie temperature, which does not allow these devices to operate at room temperature. Despite significant research efforts in the past two decades, ferromagnetism above room temperature in semiconductors remains to be elusive. A composite material consisting of ferromagnetic metal clusters embedded in a semiconducting matrix can be a good choice for spin injection.[10] Magnetic and conducting properties of such a system can also be controlled by average size and density variation of the clusters. For efficient spin injection, it is also important that the crystalline orientation between the clusters and the epitaxial semiconductor matrix is maintained everywhere. Although, the reports of semiconductor epitaxial films embedded with ferromagnetic metal clusters are not so uncommon[11–13], oriented inclusion of metallic clusters in semiconducting crystalline matrix has seldom been reported.

Nickel oxide (NiO) is a wide-bandgap (3.6 eV) semiconductor with rock-salt crystal structure.[14] The material exhibits antiferromagnetic ordering with Néel temperature of 525K.[15] While metallic nickel with fcc crystalline phase has ferromagnetic order with the ordering temperature of 631K.[16] Recently, our group have reported the growth of (111) NiO epitaxial layers embedded with (111) oriented Ni-clusters on c-sapphire substrates using pulsed laser deposition (PLD) technique. These films show ferromagnetism even at room temperature. Magnetic and conducting properties of these layers are shown to depend on the average size, aspect-ratio and density of these clusters, which can be controlled by adjusting the growth parameters.[17] Note that c-GaN epitaxial films with high crystalline quality are often grown on c-sapphire substrates. It is thus worth to explore the growth of nickel cluster embedded NiO epitaxial films on c-GaN layers. Such a composite system will be interesting not only from spin-injection perspective but also from the basic physics standpoint. For example, due to the coexistence of ferro- and antiferromagnetic phases, these layers can exhibit exchange bias effect.[18,19] Moreover, the study of magnetic anisotropy (MA) of the system could be important for data recording and storage purpose.

Here, we report the growth of (111) oriented Ni cluster embedded NiO epitaxial films on c-GaN/sapphire substrates using pulsed laser deposition technique. It has been found that the incorporation, crystalline orientation, size and density of such clusters depend strongly on the growth temperature. Cluster embedded layers are found to show ferromagnetic behaviour even at room temperature. Shape and size of the magnetic hysteresis loops are found to vary significantly with the growth temperature. This has been explained in terms of magnetic anisotropy that depends on the size, shape and orientation of the clusters.

NiO films were deposited on c-GaN/sapphire templates using pulsed laser deposition (PLD) technique, where the base pressure of the growth chamber was measured to be less than $5 \times 10^{-6}$ mbar. A KrF excimer laser with wavelength of 248 nm and pulse width of 25 ns was used to ablate the NiO pellet. Energy density of the laser pulse was kept at 1.5 Jcm$^{-2}$ at a frequency of 5 Hz. Several samples were grown at different growth temperatures ($T_G$) and oxygen pressures ($P_{O_2}$). Growth time was adjusted to 5000 pulses that corresponds to 16.67 min for all samples. Both in- and out-of-plane x-ray diffraction studies were carried out on these samples using a Rigaku Smart Lab high-resolution X-ray diffraction (HR-XRD) system. Surface morphology was investigated by atomic force microscopy (AFM) and field emission scanning electron microscopy (FE-SEM). For microstructural (crystal mapping) studies precession electron diffraction (PED) technique was used from NanoMegas ASTAR® (Brussels, Belgium) with a spatial resolution of 1 nm and specimen preparation were carried out using focussed ion beam (FIB) lift-out techniques using a Helios 5 UC Thermoscientific dual-beam instrument with a resolution ~0.6 nm. Magnetization measurements were carried out using a superconducting quantum interference device (SQUID) vibrating sample magnetometer (VSM) for fields ranging between ±4T applied parallel and perpendicular to the sample surface and at temperatures ranging from 100 to 300K. All data presented here were corrected for the diamagnetic background of the substrate.

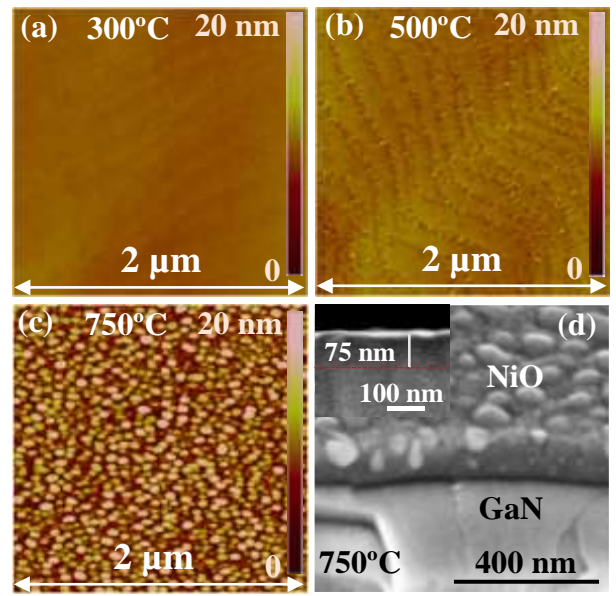

**Fig.1:** AFM surface images for the NiO films grown at (a) 300ºC, (b) 500ºC and (c) 750ºC. (d) high-resolution SEM surface image for the 750ºC sample recorded at 45º angle with respect to the surface. Inset shows the cross-sectional image for the same sample.

Several samples were grown at a growth temperature of 500ºC under various oxygen pressures ranging from $1 \times 10^{-2}$ to $1 \times 10^{-5}$ mbar. Detailed XRD, AFM and FE-SEM studies of these samples suggest that the incorporation of Ni clusters in NiO matrix takes place only when the pressure of oxygen is sufficiently low. These results are discussed in the supplementary figure S-1 and S-2. Here, we focus on the set of samples, which are grown at various growth temperatures ranging from 300 to 750ºC, while the oxygen pressure during growth is maintained at $1 \times 10^{-5}$ mbar in all cases.

Fig.1(a), (b) and (c) show the AFM images for the NiO films grown at different temperatures. While the 300 and 500ºC samples show smooth and continuous morphology, a large density of circular mound structures could be seen on the surface in case of the 750ºC sample. Fig.1(d) presents the high-resolution SEM surface image for the 750ºC sample recorded at an angle of 45º with respect to the surface. A large density of clusters, which are embedded in the NiO matrix, is clearly visible in the picture. It seems that the top of each cluster protrudes as mound on the surface. Inset shows the cross-sectional SEM image for the sample. Deposition of a continuous film of thickness 75 nm is evident from the image. All samples investigated here are found to be almost of the same thickness.

Fig.2(a) shows the out-of-plane XRD $\omega - 2\theta$ scans recorded for samples grown at different temperatures. Apart from (111) and its higher order reflections, no other planes of NiO are visible in these scans. This suggests [111] oriented growth of NiO. Interestingly, in all the samples, an additional feature appears at around $2\theta = 44.5º$. Intensity of this peak increases with $T_G$. This reflection can be attributed to (111) plane associated with the fcc crystalline phase of Ni. The finding thus implies the inclusion of Ni clusters in NiO layer with Ni(111)∥NiO(111) orientation. Further, the inclusion increases with the growth temperature. Fig. 2(b) presents the in-plane $2\theta_\chi - \phi$ scans for these samples. $(2\bar{2}0)$ is the only reflection visible from NiO phase along with the $(11\bar{2}0)$ peak of GaN and $(2\bar{2}0)$ reflection of Ni. Note that Ni(111) reflection is also visible in the $2\theta_\chi - \phi$ scans of these samples, which suggests that all the Ni clusters are not oriented with Ni(111) ∥ NiO(111). Few of the clusters have their [111]-direction lying parallel to the sample surface (L-clusters). In-plane reciprocal space map (RSM) for the sample grown at 500°C is shown in the supplementary figure S-3. This finding clearly demonstrates that NiO($1\bar{1}0$) ∥ GaN($11\bar{2}0$) in-plane and NiO(111) ∥ GaN(0001) out-of-plane orientation relationships are maintained between NiO and GaN. Moreover, most of the Ni clusters are also crystallographically oriented following Ni(111) ∥ NiO(111) and Ni($1\bar{1}0$) ∥ NiO($1\bar{1}0$) relationship. Interestingly, the findings of the supplementary figure S-3 also implies that Ni(111) ∥ GaN($11\bar{2}0$) relationship is obeyed for the L-clusters.

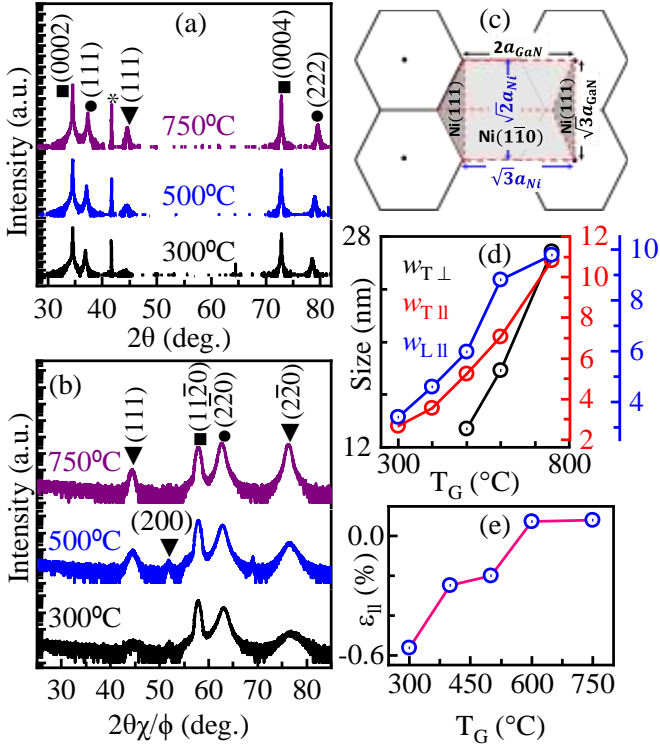

Fig.2: High-resolution (a) ω-2θ and (b) 2θχ-φ profiles for samples grown at different growth temperature. In both the panels, solid squares, circles, triangles and stars represent the reflections from GaN, NiO, Ni and sapphire, respectively. (c) Schematic presentation of the basal plane orientation of L-type nickel clusters with respect to (0001) plane of GaN. (d) Spatial extent of the T- and L-clusters in (∥) and out (⊥) of the sample plane as functions of the growth temperature. (e) Variation of the biaxial strain ($\varepsilon_\parallel$) of the T-clusters with the growth temperature.

Note that three times the separation between the (111) planes of Ni lattice ($= 3 \times a_{Ni}/\sqrt{3}$, where $a_{Ni} = 3.529$Å, the lattice constant of Ni) comes out quite close to two times the $a$-lattice constant of GaN ($a_{GaN} = 3.153$Å) with only 3% mismatch. Similarly, the base length of (111) triangular planes of Ni ($= \sqrt{2}a_{Ni}$) is close to the separation between the centres of two adjacent hexagons of GaN lattice ($= \sqrt{3}a_{GaN}$) with only 9% mismatch as shown schematically in Fig. 2(c). Therefore, the inclusion of Ni-clusters with [111]-direction lying in the plane of the sample is in principle possible. This speculation is also consistent with the Ni(111) ∥ GaN($11\bar{2}0$) relationship observed for the L-clusters (see supplementary figure S-3). A tiny hump could also be seen at $2\theta_\chi \approx 52°$ in $2\theta_\chi - \phi$ scans for samples grown at $T_G < 750$°C. This peak can be attributed to Ni(200) reflection stemming from Ni clusters with other crystalline orientations.

Average spatial extents in the out-of-plane ($w_{T\perp}$) and in-plane ($w_{T\parallel}$) directions of the clusters, whose [111] direction is perpendicular to the sample surface (T-clusters), have been estimated from the widths of the Ni(111) peak in $\omega - 2\theta$ scans and Ni($2\bar{2}0$) peak in $2\theta_\chi - \phi$ scans, respectively, using Scherrer formula. L-clusters can only be evidenced through the Ni(111) peak in $2\theta_\chi - \phi$ scans. The average size of these clusters in the in-plane directions ($w_{L\parallel}$) can be estimated from the width of the Ni peak. Fig. 2(d) shows the variation of $w_{T\perp}$, $w_{T\parallel}$ and $w_{L\parallel}$ as functions of the growth temperature. It is evident that the overall size of both the cluster types increases with $T_G$. Note that as compared to other samples, the intensity ratio between Ni($2\bar{2}0$) and Ni(111) peaks in the $2\theta_\chi - \phi$ scans is significantly high for the sample grown at 750°C implying that the sample has the highest proportion of T-clusters (see supplementary figure S-4). A careful examination of the $\omega - 2\theta$ and $2\theta_\chi - \phi$ profiles reveal that T-clusters are biaxially strained. The in-plane strain $\varepsilon_\parallel$ is plotted as a function of the growth temperature for these samples in Fig.2(e). Clearly, almost 0.6% compressive strain exists in the clusters at $T_G = 300$°C. The strain decreases as $T_G$ is increased. This is also consistent with the size variation of the clusters with $T_G$.

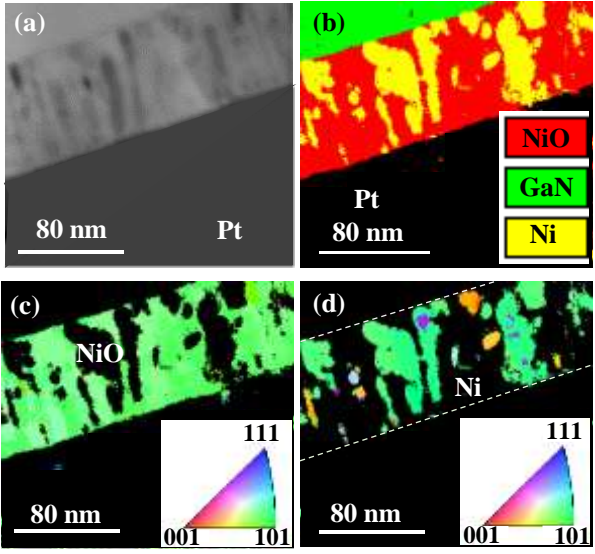

**Fig.3:** Precession electron diffraction (a) virtual bright field (VBF) map of cross-sectional lamella, (b) phase map, (c) orientation map for the NiO matrix and (d) orientation map for Ni clusters in case of the sample grown at 600ºC.

Fig.3 presents the PED results obtained by cross-sectional transmission electron microscopy (TEM) on the lamella prepared from 600ºC grown sample using the FIB technique. Fig. 3(a) shows the virtual bright field (VBF) cross-sectional map of the lamella. Inclusion of clusters, which are elongated along the perpendicular direction to the surface, is quite evident from the image. Fig.3(b) and (c) show the phase map and the orientation map for NiO, which clearly demonstrates the existence of Ni-clusters in the NiO matrix. The orientation map also confirms the epitaxial nature of NiO film. Fig.3(d) depicts the orientation map for Ni-clusters. It is quite evident from the figure that most of the clusters are crystallographically aligned with respect to the NiO matrix in the in-plane direction, which is consistent with the observations of XRD study shown in Fig.2. Moreover, the shape and size of these clusters are also in agreement with the XRD results. Fig. 3(d) also shows the presence of Ni-clusters with other crystalline orientations including L-clusters (purple regions).

Fig.4(a) and (b) show the magnetization loops (symbols) recorded at 100K for samples grown at 500ºC with the magnetic field applied in the plane and perpendicular to the plane of the sample, respectively. Evidently, in both the cases, the loops are cramped at the middle, which indicates the coexistence of two ferromagnetic phases with different coercivities. The magnetization $M$ versus the field $H$ loops can be deconvoluted using the following equation[20]

$$M(H) = \sum_{i=1}^{2} \frac{2M_S^i}{\pi} \tan^{-1}\left[\frac{(H \pm H_C^i)}{H_C^i} \tan\left(\frac{\pi S^i}{2}\right)\right] \quad (1)$$

where, $H_C^i$ and $M_S^i$ are the coercivity and saturation magnetization, while $S^i$ represents the ratio between the remnant and saturation magnetization ($M_R^i/M_S^i$) for the i-th magnetic phase. Deconvoluted loops along with the fitting are also shown in the respective figures. The loop with larger saturation magnetization can be attributed to the T-clusters as they dominate over the clusters with other orientations (see Fig. 2 and 3). While, the loop with lower magnetic saturation must be associated with other clusters such as L-clusters.

Fig.4(c) compares the magnetization loops resulting only from the T-clusters at 100K for the magnetic field applied along (IP) and perpendicular (OP) to the sample plane in case of the 500ºC grown sample. Clearly, the coercive field ($H_c$) as well as $M_r/M_s$ ratio are more in the in-plane case implying that the easy-axis of the T-clusters is lying in the in-plane direction for the sample. However, the magnetization loops recorded at 300K for this sample show higher value of $H_c$ in the out-of-plane direction (see supplementary figure S-5). Similar findings have been obtained in case of sample grown at 400ºC. This may indicate a reversal of easy-axis at 300K in these samples. Fig.4(d) compares the magnetization loops recorded at 100K with IP and OP directional magnetic fields for the sample grown at 750ºC. Shape of the loops clearly suggests the presence of a single ferromagnetic phase in this sample. This is consistent with the fact that at $T_G =750$ºC, T-clusters dominate by a big margin over other clusters (see supplementary figure S-4). Clearly, in this sample, both $H_c$ and $M_r/M_s$ ratio are higher for OP case, suggesting that the T-clusters in this sample have out-of-plane anisotropy. Magnetization loops recorded at 300K for this sample also show the out-of-plane alignment of the easy-axis (see supplementary figure S-5), meaning no reversal of anisotropy with increasing temperature. Similar observations can be found in case of sample grown at 600ºC.

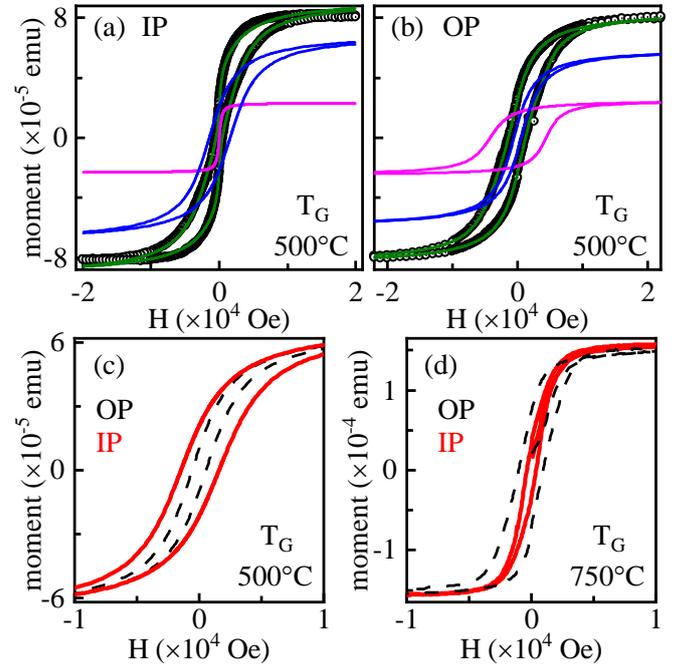

**Fig.4:** Magnetization loops (symbols) recorded at 100K for samples grown at 500ºC for the magnetic field applied (a) in the plane (IP) and (b) perpendicular to the plane (OP) of the sample surface. Deconvolution of the profiles using eq. (1) is also shown in respective figures. Magnetization loops for IP and OP orientations of the magnetic field in case of (c) the deconvoluted contribution only from the T-clusters in 500ºC sample and (d) the sample grown at 750ºC.

Magnetocrystalline anisotropy (McA) should favor [111] direction as the easy-axis in bulk fcc phase of nickel. However, for such small size Ni-clusters (few tens of nanometers) found in these samples, other anisotropy contributions, such as shape anisotropy (ShA), magnetoelastic anisotropy (MeA) and surface anisotropy (SuA), should play more important roles than McA. Note that the T-clusters are found to be elongated along [111]-direction. ShA should thus favor uniaxial anisotropy along [111]-direction for these clusters. It should be noted that the average size of the T-clusters increases with the growth temperature [see Fig. 2(d)]. Surface anisotropy energy per unit volume $\eta_{SuA}$, which should scale with the inverse of the size of the clusters[21], is thus expected to increase with the decrease of $T_G$. The origin of SuA for these clusters could be the exchange coupling between the ferromagnetic clusters and the antiferromagnetic NiO matrix. Another possible reason could be the anisotropic occupancy of Ni 3$d$ orbital in NiO matrix at the interface.[22] Moreover, the biaxial strain in these clusters has been found to increase with the reduction of $T_G$ as shown in Fig. 2(e). This could mean that the magnetoelastic anisotropy energy per unit volume $\eta_{MeA}$ should also increase as the growth temperature is reduced. Both $\eta_{SuA}$ and $\eta_{MeA}$ can, in principle, compete with the shape anisotropy $\eta_{ShA}$ specially in samples grown at lower growth temperatures. Size dependence of anisotropy has indeed been reported in various types of magnetic systems and is often attributed to surface and magnetoelastic effects[22–24]. We believe that the reason for in-plane anisotropy observed at 100K for the T-clusters in samples grown at 400 and 500°C is the domination of either MeA or SuA (or both) over the shape anisotropy. This competition might be further reflected in the observation of temperature dependent reversal of easy-axis in samples grown at lower temperatures (400 and 500°C). It is plausible that the strength of the exchange coupling between the Ni-ions at the cluster/matrix interface, which provides the surface anisotropy contribution to the clusters, diminishes at a faster rate with increasing temperature for smaller clusters.

In conclusions, (111) NiO epitaxial films embedded with Ni clusters can be grown on c-GaN/sapphire substrates using pulsed laser deposition technique. Most of the clusters are found to be crystallographically aligned with the NiO matrix. However, clusters with other orientations also exist, specially at lower growth temperatures. Growth temperature is found to play a crucial role in governing the inclusion, orientation, shape, size and density of these clusters. Proportion as well as size of clusters with Ni(111)∥NiO(111) orientation increases with the growth-temperature. Cluster embedded layers exhibit ferromagnetic behaviour even at room temperature. Magnetic anisotropy of these clusters is also found to depend strongly on the growth temperature. While the anisotropy is found to be in-planer for samples grown at lower temperatures. Easy-axis turns perpendicular to the layer plane for the samples grown at higher temperatures. This has been explained in terms of a cluster size dependent trilateral competition between the shape, magnetoelastic and surface anisotropies.

The authors acknowledge financial support from the Department of Science and Technology (DST), Government of India, under Grant No: CRG/2018/001343. They would also like to acknowledge the use of various facilities under the Industrial Research and Consultancy Centre (IRCC), Sophisticated Analytical Instrument Facility (SAIF) and the Centre for Excellence in Nanoelectronics (CEN), IIT Bombay and Professor K.G. Suresh, Department of Physics, IIT Bombay, for valuable discussions.

**References**

[1] T.J. Flack, B.N. Pushpakaran, and S.B. Bayne, J. Electron. Mater. **45**, 2673 (2016).

[2] G. Zulauf, S. Park, W. Liang, K.N. Surakitbovorn, and J. Rivas-Davila, IEEE Trans. Power Electron. **33**, 10748 (2018).

[3] Y. Zhong, J. Zhang, S. Wu, L. Jia, X. Yang, Y. Liu, Y. Zhang, and Q. Sun, Fundam. Res. **2**, 462 (2022).

[4] H. Amano, Y. Baines, E. Beam, M. Borga, T. Bouchet, P.R. Chalker, M. Charles, K.J. Chen, N. Chowdhury, R. Chu, C. De Santi, M.M. De Souza, S. Decoutere, L. Di Cioccio, B. Eckardt, T. Egawa, P. Fay, J.J. Freedsman, L. Guido, O. Häberlen, G. Haynes, T. Heckel, D. Hemakumara, P. Houston, J. Hu, M. Hua, Q. Huang, A. Huang, S. Jiang, H. Kawai, D. Kinzer, M. Kuball, A. Kumar, K.B. Lee, X. Li, D. Marcon, M. März, R. McCarthy, G. Meneghesso, M. Meneghini, E. Morvan, A. Nakajima, E.M.S. Narayanan, S. Oliver, T. Palacios, D. Piedra, M. Plissonnier, R. Reddy, M. Sun, I. Thayne, A. Torres, N. Trivellin, V. Unni, M.J. Uren, M. Van Hove, D.J. Wallis, J. Wang, J. Xie, S. Yagi, S. Yang, C. Youtsey, R. Yu, E. Zanoni, S. Zeltner, and Y. Zhang, J. Phys. D. Appl. Phys. **51**, 163001 (2018).

[5] S. Krishnamurthy, M. van Schilfgaarde, and N. Newman, Appl. Phys. Lett. **83**, 1761 (2003).

[6] S.P. Dash, S. Sharma, R.S. Patel, M.P. de Jong, and R. Jansen, Nature **462**, 491 (2009).

[7] Q. Wu, D. Lin, M. Chen, J. Li, W. Hu, X. Wu, F. Xu, C. Zhang, Y. Cao, X. Li, Y. Wu, Z. Wu, and J. Kang, Appl. Phys. Lett. **122**, (2023).

[8] M. C. Prestgard, G. P. Siegel, and A. Tiwari, Adv. Mater. Lett. **5**, 242 (2014).

[9] I. Žutić, J. Fabian, and S. Das Sarma, Rev. Mod. Phys. **76**, 323 (2004).

[10] M. Jamet, A. Barski, T. Devillers, V. Poydenot, R. Dujardin, P. Bayle-Guillemaud, J. Rothman, E. Bellet-Amalric, A. Marty, J. Cibert, R. Mattana, and S. Tatarenko, Nat. Mater. **5**, 653 (2006).

[11] A. V. Kudrin, V.P. Lesnikov, Y.A. Danilov, M. V. Dorokhin, O. V. Vikhrova, P.B. Demina, D.A. Pavlov, Y. V. Usov, V.E. Milin, Y.M. Kuznetsov, R.N. Kriukov, A.A. Konakov, and N.Y. Tabachkova, Semicond. Sci. Technol. **35**, 125032 (2020).

[12] H. Li, C. Wang, D. Li, P. Homm, M. Menghini, J.-P. Locquet, C. Van Haesendonck, M.J. Van Bael, S. Ruan, and Y.-J. Zeng, J. Phys. Condens. Matter **31**, 155301 (2019).

[13] S. Dhar, O. Brandt, A. Trampert, L. Däweritz, K.J. Friedland, K.H. Ploog, J. Keller, B. Beschoten, and G. Güntherodt, Appl. Phys. Lett. **82**, 2077 (2003).

[14] W. Chia-Ching and Y. Cheng-Fu, Nanoscale Res. Lett. **8**, 33 (2013).

[15] T.M. Schuler, D.L. Ederer, S. Itza-Ortiz, G.T. Woods, T.A. Callcott, and J.C. Woicik, Phys. Rev. B **71**, 115113 (2005).

[16] B. Legendre and M. Sghaier, J. Therm. Anal. Calorim. **105**, 141 (2011).

[17] S.K. Yadav, B.P. Sahu, and S. Dhar, J. Phys. D. Appl. Phys. **55**, 035002 (2022).

[18] T. Blachowicz and A. Ehrmann, Coatings **11**, 122 (2021).

[19] X.-J. Yao, X.-M. He, X.-Y. Song, Q. Ding, Z.-W. Li, W. Zhong, C.-T. Au, and Y.-W. Du, Phys. Chem. Chem. Phys. **16**, 6925 (2014).

[20] M.B. Stearns and Y. Cheng, J. Appl. Phys. **75**, 6894 (1994).

[21] D.A. Garanin and H. Kachkachi, Phys. Rev. Lett. **90**, 065504 (2003).

[22] Y.-J. Zhang, L. Wu, J. Ma, Q.-H. Zhang, A. Fujimori, J. Ma, Y.-H. Lin, C.-W. Nan, and N.-X. Sun, Npj Quantum Mater. **2**, 17 (2017).

[23] S.E. Shirsath, X. Liu, Y. Yasukawa, S. Li, and A. Morisako, Sci. Rep. **6**, 30074 (2016).

[24] Z. Ma, J. Mohapatra, K. Wei, J.P. Liu, and S. Sun, Chem. Rev. **123**, 3904 (2023).

Supplementary Material

# Ni cluster embedded (111)NiO layers grown on (0001)GaN films using pulsed laser deposition technique

**Simran Arora[1], Shivesh Yadav[2], Amandeep Kaur[1], Bhabani Prasad Sahu[1], Zaineb Hussain[1], Subhabrata Dhar[1]***

1) Department of Physics, Indian Institute of Technology Bombay, Mumbai-400076, India

2) Department of Condensed Matter Physics and Material Science, Tata Institute of Fundamental Research, Mumbai – 400005, India

*E-mail: dhar@phy.iitb.ac.in

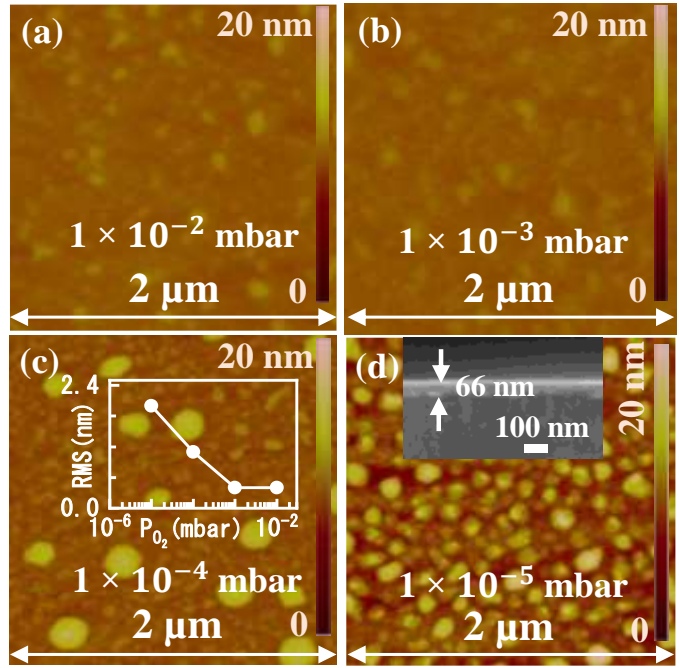

**Fig.S-1:** AFM surface images of NiO films grown at the oxygen pressure of (a) 1×10⁻² (b) 1×10⁻³ (c) 1×10⁻⁴ and (d) 1×10⁻⁵ mbar inside the chamber. Inset of (c) shows the RMS roughness of films as a function of the oxygen pressure and the inset of (d) shows the cross-sectional SEM image of the sample, which shows the growth of continuous and smooth film of thickness ~ 60 nm. The density of circular mounds on the surface could be seen to increase with the reduction of the oxygen pressure

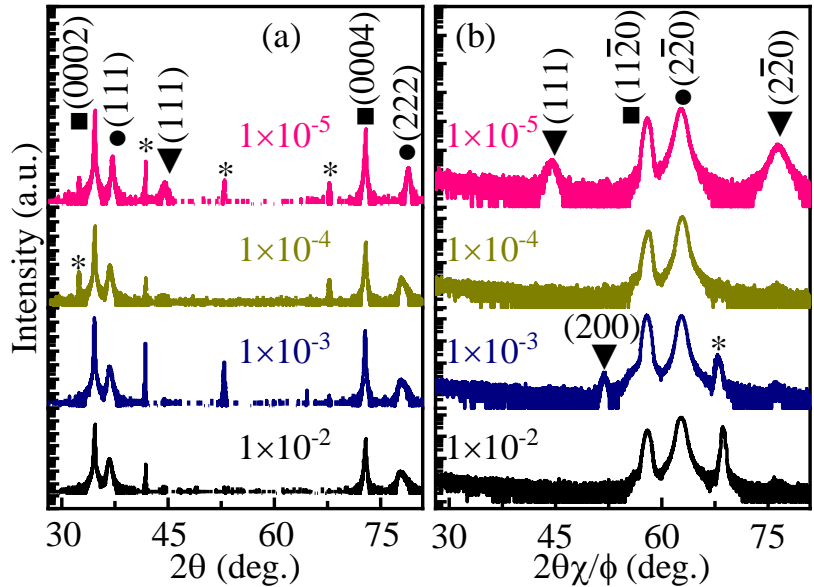

**Fig.S-2:** High-resolution X-ray diffraction (a) ω-2θ scans (b) 2θχ-φ profiles for samples grown at different oxygen pressure (In both the panels, solid squares, circles, triangles and stars represent the reflections from GaN, NiO, Ni and sapphire, respectively). Reflections from crystalline planes of Ni could only be seen in sample grown at 1×10⁻⁵ mbar of oxygen pressure.

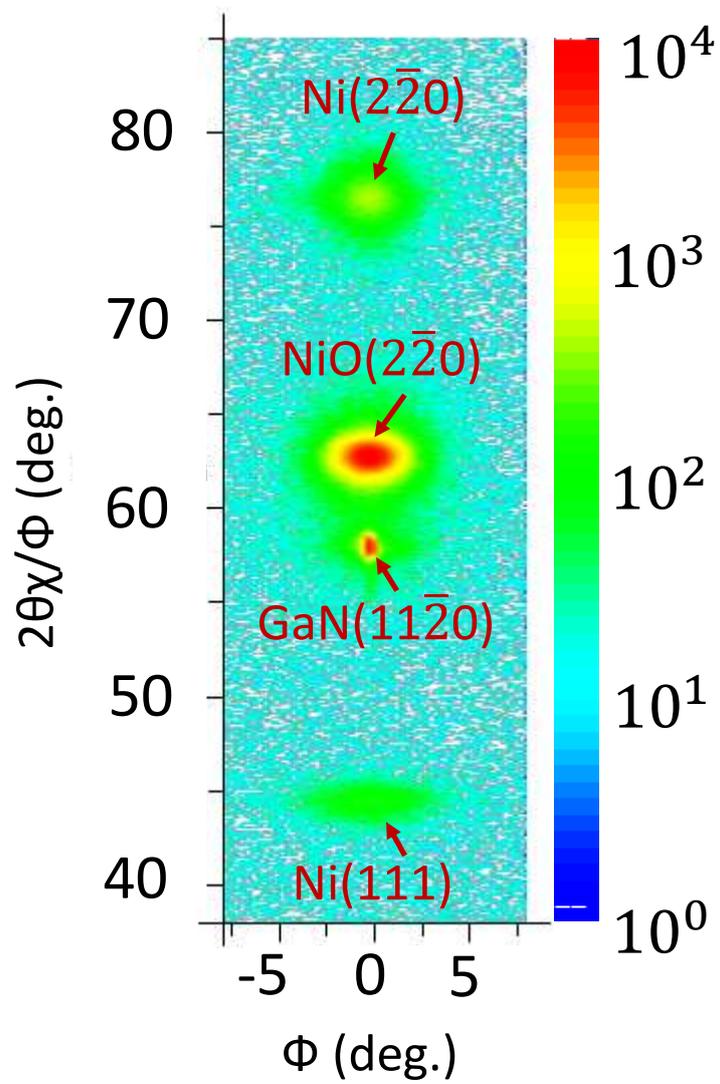

**Fig.S-3:** In-plane reciprocal space map (RSM) for the sample grown at 500ºC under $1\times10^{-5}$ mbar oxygen pressure. GaN (11$\bar{2}$0), NiO (2$\bar{2}$0), Ni (2$\bar{2}$0) [arising from the T-clusters] and Ni (111) [arising from the L-clusters] reflections. This finding clearly demonstrates that NiO(2$\bar{2}$0) ∥ GaN(11$\bar{2}$0) in-plane orientation relationship. Furthermore, T-type of Ni clusters also crystallographically oriented with the NiO matrix satisfying Ni(2$\bar{2}$0) ∥ NiO(2$\bar{2}$0) relationship. L-type of clusters follow Ni(111)∥GaN(11$\bar{2}$0) relationship as predicted in Fig. 2(c) of the manuscript.

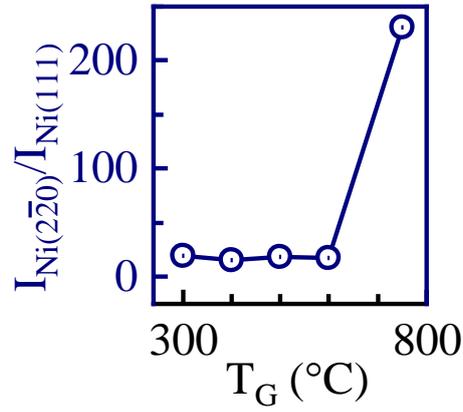

**Fig.S-4:** Structure-factor renormalized intensity ratio between the Ni(2$\bar{2}$0) [attributed to the T-clusters] to Ni(111) [attributed to the L-clusters] reflections obtained from the $2\theta_\chi - \phi$ profiles as a function of the growth temperature. This quantity can serve as a good measure for the ratio of the proportions of the two types of clusters in the film. Interestingly, the ratio shows an order of magnitude enhancement as $T_G$ increases from 600 to 750ºC.

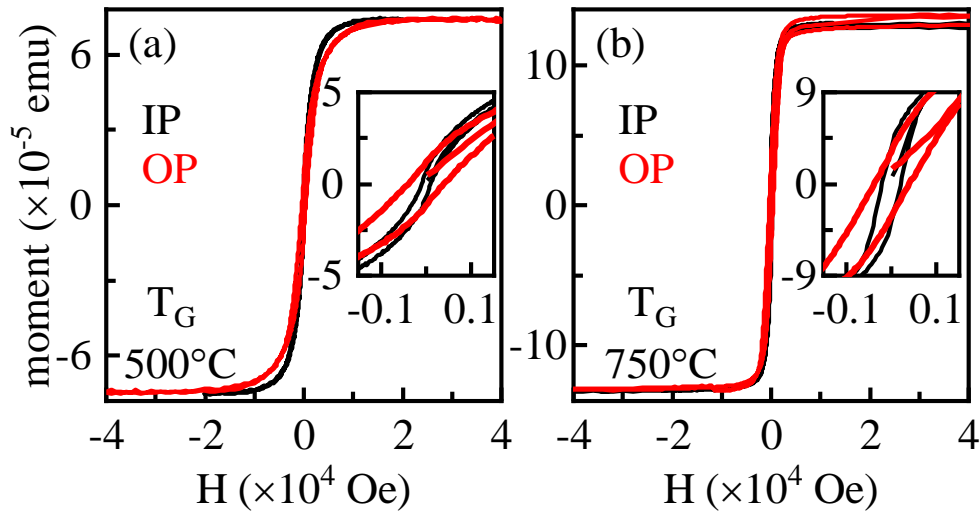

**Fig.S-5:** Magnetization loops recorded at 300K for samples grown at (a) 500ºC and (b) 750ºC for the magnetic field applied in the plane (IP) and perpendicular to the plane (OP) of the sample surface. Respective insets show the loops in expanded scales. Higher value of the coercive field ($H_c$) in OP as compared to IP configuration in the sample grown at 500ºC is noticeable. Note that 100K data for the sample show higher value of $H_c$ in IP configuration [see Fig. 4(c) of the manuscript]. This may suggest a reversal of easy-axis at 300K. However, $H_c$ remains to be more in OP configuration for the sample grown at 750ºC even at 300K implying no reversal of easy-axis.